\documentstyle[prl,aps]{revtex}
\input BoxedEPS.tex
\SetOzTeXEPSFSpecial
\HideDisplacementBoxes

\begin{document}

\twocolumn[\hsize\textwidth\columnwidth\hsize
           \csname @twocolumnfalse\endcsname
\title{Argon annealing of the oxygen-isotope
exchanged manganite La$_{0.8}$Ca$_{0.2}$MnO$_{3+y}$}

\author{Guo-meng~ Zhao, K. Conder, H.~ Keller and K. A. M\"uller}
\address{Physik-Institut der Universit\"at Z\"urich,
CH-8057 Z\"urich, Switzerland}

\maketitle
\widetext

\begin{abstract}

We have resolved a controversial issue concerning the oxygen-isotope
shift of the ferromagnetic transition temperature T$_{C}$ in
the manganite La$_{0.8}$Ca$_{0.2}$MnO$_{3+y}$. We show that the giant
oxygen-isotope shift of T$_{C}$ observed in the normal oxygen-isotope
exchanged samples is indeed intrinsic, while a much smaller shift observed in
the argon annealed samples is an artifact. The argon annealing causes the
$^{18}$O sample to
partially exchange back to the $^{16}$O isotope due to a small $^{16}$O$_{2}$
contamination in the Ar gas. Such a contamination is commonly caused by
the oxygen outgas that is trapped
in the tubes, connectors and valves. The present
results thus umambiguously demonstrate that the observed large oxygen
isotope effect is an intrinsic property of manganites, and places an important
constraint on the basic physics of these materials.

\end{abstract}
\vspace{0.5cm}
\narrowtext
]
An intensive research effort \cite{Art} has recently been made to study the
manganese-based perovskites Ln$_{1-x}$A$_{x}$MnO$_{3}$ (where Ln is a
trivalent element, A is
a divalent element) due to the discovery of very
large (``colossal'') magnetoresistance (CMR) in thin films of these
compounds \cite{Von,Jin}. The physics of manganites has primarily been
described by the
double-exchange model \cite{Zener,Anderson}. Recent calculations
\cite{Millis1,Roder,Alex} show that a strong electron-phonon
interaction must be involved to explain the basic physics of these
materials. Many
recent experiments have provided compelling evidence for the existence
of a strong
electron-phonon interaction and of polaronic charge carriers in manganites
\cite{Jaime,ZhaoNature,ZhaoPRL,Kim,ZhaoJPCM,Billinge,Teresa,Booth,Worledge}.

In particular, the observed giant oxygen isotope shift of the Curie
temperature \cite{ZhaoNature}
should provide direct evidence that lattice vibrations play an important
role in the magnetic properties of these materials. However,
Nagaev \cite{Nagaev} has recently shown that the
observed giant oxygen-isotope effects in manganites \cite{ZhaoNature} are not
caused by a strong electron-phonon coupling, but rather by an oxygen-mass
dependence of excess oxygen. In addition, Franck {\em et al.,}
demonstrated that \cite{Franck98} the oxygen-isotope shift
was reduced by more than 10 K after the oxygen-isotope exchanged
samples of La$_{0.8}$Ca$_{0.2}$MnO$_{3+y}$ were annealed for 24 h in argon and
at 950 $^{\circ}$C. They thus argued that the giant isotope shift (21
K) reported in Ref. \cite{ZhaoNature} is not intrinsic, but caused by
the presence of excess oxygen in the samples.

In order to resolve the controvercy concerning the isotope effect in
the manganites, we perform thermogravimetry (TG) experiments on
the oxygen-isotope exchanged samples of La$_{0.8}$Ca$_{0.2}$MnO$_{3}$,
which were carried
out in flowing argon
gas and at 950 $^{\circ}$C. The experiments demonstrate that the $^{18}$O
sample was partially exchanged back to the $^{16}$O isotope when it was
annealed in flowing argon gas and at 950 $^{\circ}$C.
This is due to the fact that the oxygen outgas trapped in
the tubes, connectors and valves contaminates the argon gas although
the Ar gas itself is very pure. The present
experiments thus show that the
oxygen-isotope effects observed in the argon annealed samples are not
reliable, and that the normal isotope exchange procedure can ensure 
the same oxygen content for two isotope samples.

Samples of La$_{0.8}$Ca$_{0.2}$MnO$_{3+y}$ were
prepared by
conventional solid state reaction using dried La$_{2}$O$_{3}$, MnO$_{2}$
and CaCO$_{3}$.  The well-ground mixture was
heated in
air at 1000 $^{\circ}$C for 20 h, 1100 $^{\circ}$C for 20 h
with one intermediate grinding. The powder
samples were then pressed into pellets
and sintered at 1260 $^{\circ}$C for 72 h, and 1160 $^{\circ}$C for
72 h with one intermediate grinding. Two  pieces were
cut
from the same pellet for oxygen-isotope diffusion. The diffusion was
carried out
for 50 h
at 1000 $^{\circ}$C and oxygen pressure of 1 bar.
The oxygen-isotope enrichment was determined from the weight changes
of both $^{16}$O and $^{18}$O samples. The $^{18}$O samples had
$\sim$90$\%$
$^{18}$O
and $\sim$10$\%$ $^{16}$O.

Thermogravimetry (TG) experiments were performed using PERKIN ELMER
TGA7 Instrument. The investigated samples were heated in a stream
(50 cm$^{3}$/min) of very pure Ar (99.998$\%$). The weights of
the $^{16}$O and $^{18}$O samples used for TG
experiments were 71.833 mg and 46.145 mg, respectively.
Before each experiment, the
balance (with the sample inside) was flushed with the pure Ar at
room temperature for at least 40 h.

Fig. 1 shows the TG data for both $^{16}$O and
$^{18}$O samples of La$_{0.8}$Ca$_{0.2}$MnO$_{3}$. The weight was
renormalized to that at 500
$^{\circ}$C to eliminate the error due to adsorption of water and
CO$_{2}$ in the samples. From the figure, one can see that the weights of
both isotope samples start to decrease
when the temperature reaches 950 $^{\circ}$C. However, there is
a substantial difference in the weight loss for the two isotope samples.
After argon annealing at 950 $^{\circ}$C for 150 minutes, the weight
of the $^{16}$O sample decreases by about 0.20$\%$ while the weight of
the $^{18}$O is reduced by 1.60$\%$. After argon annealing at 950
$^{\circ}$C for 24 hours, the weight
of the $^{16}$O sample decreases by about 0.25$\%$ while the weight of
the $^{18}$O is reduced by 2.50$\%$. The extra weight decrease for
the $^{18}$O sample is due to the fact that the $^{18}$O sample was
partially exchanged back to the $^{16}$O isotope because of the $^{16}$O
contamination in the Ar gas. Such a contamination is commonly
caused by the oxygen outgas that was trapped
in the tubes, connectors and valves. Without heating these elements
in the system, it is hard to get rid of the trapped outgas, and the
contamination is unavoidable. From the weight changes, we can
estimate that the $^{18}$O content of the $^{18}$O sample became
about 40$\%$ after annealing for 150 minutes, and about 5$\%$ after
annealing for 24 h.
\begin{figure}[htb]
    \ForceWidth{7.5cm}
	\centerline{\BoxedEPSF{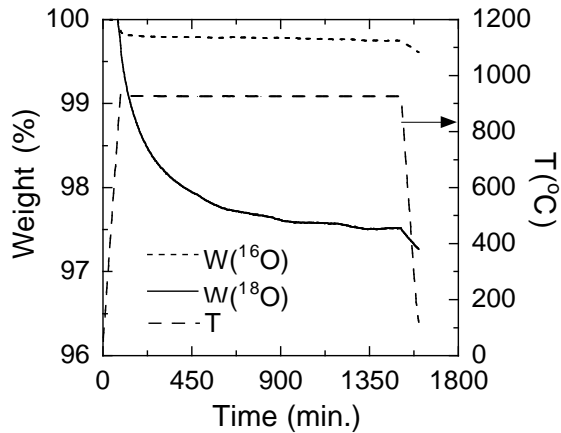}}
	\vspace{0.5cm}
	\caption[~]{Thermogravimetry (TG) data for both $^{16}$O and
$^{18}$O samples of La$_{0.8}$Ca$_{0.2}$MnO$_{3}$. The
short-dash line and solid line are for the weights of the $^{16}$O and
$^{18}$O samples (left scale), respectively. The long-dash line is for
the temperature profile of both isotope samples (right scale). The
investigated samples were heated in a stream
(50 cm$^{3}$/min) of pure Ar (99.998$\%$). The weights of the $^{16}$O and
$^{18}$O samples used for TG
experiments were 71.833 mg and 46.145 mg, respectively.
Before each experiment, the
balance (with the sample inside) was flushed with the pure Ar at
room temperature for at least 40 h. }
	\protect\label{Fig.1}
\end{figure}

In order to check the influence of the argon annealing on the
ferromagnetic transition temperature T$_{C}$, we performed magnetization
measurements for these samples.  Field-cooled magnetization was
measured with a Quantum Designed SQUID magnetometer in a field of  5 mT.
The samples were cooled
directly to 5 K, then warmed up to a temperature well below T$_{C}$.
After waiting for 5 minutes at that temperature, data were
collected upon warming to a temperature well above T$_{C}$. In Fig.~2, we
plot the temperature dependence of the low-field
magnetization (normalized to the magnetization
well below T$_{C}$) for the $^{16}$O and $^{18}$O samples
of La$_{0.8}$Ca$_{0.2}$MnO$_{3}$ (a) before the argon annealing; (b)
after the argon annealing at 950 $^{\circ}$C for 24 h. It is clear that
before the argon annealing, the oxygen-isotope shift of T$_{C}$ is about
21 K, while the shift becomes very small (about 1 K) after the annealing.
As shown above, the $^{18}$O sample contains only
about 5$\%$ $^{18}$O, so the isotope shift should be about 1 K, as
observed. The
result clearly shows that a very small isotope shift observed in the
present argon
annealed samples is due to a small $^{16}$O contamination in the argon
gas, which is sufficient to exchange the $^{18}$O back to the
$^{16}$O isotope.
\begin{figure}[htb]
    \ForceWidth{7.5cm}
	\centerline{\BoxedEPSF{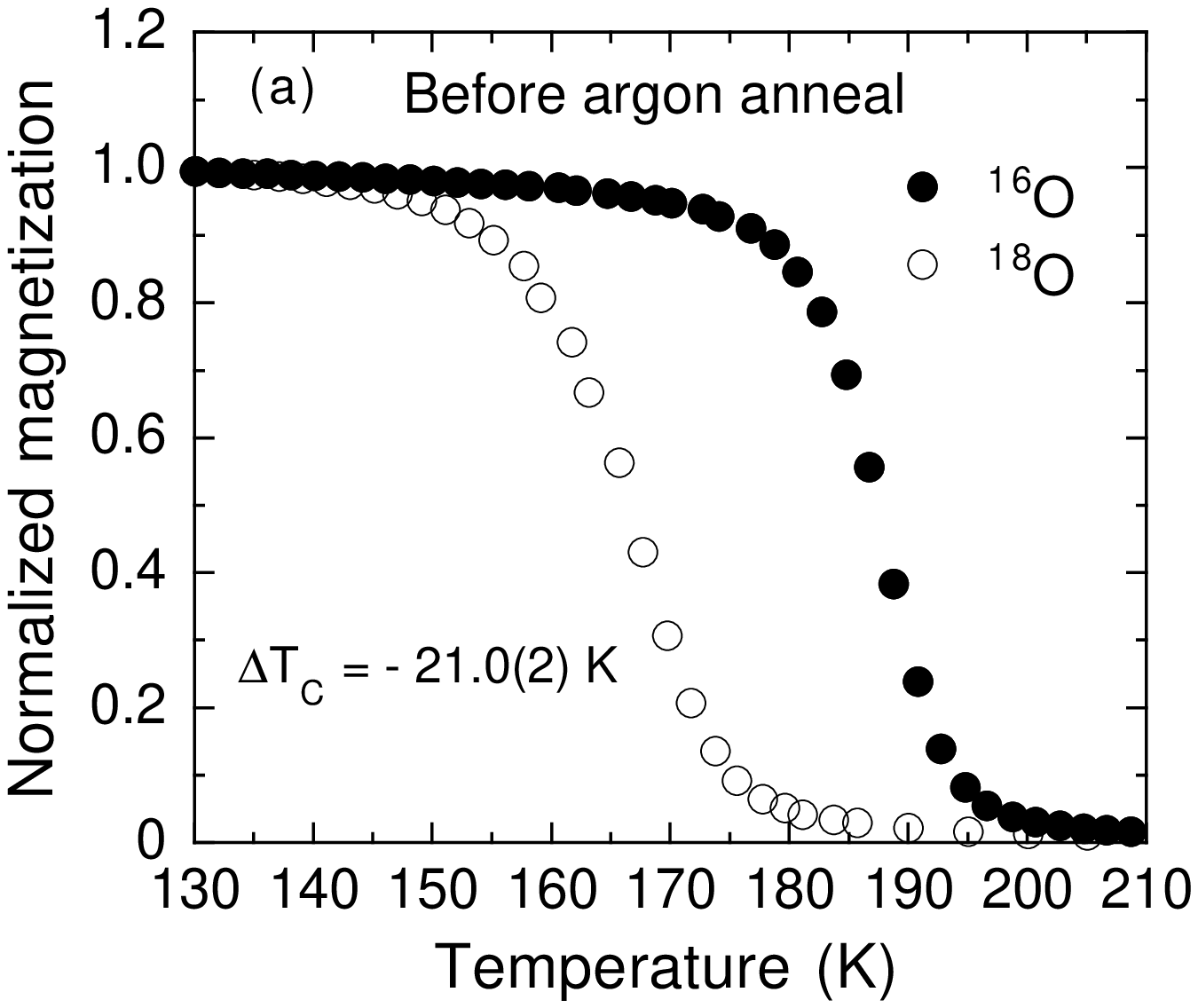}}
	\vspace{0.5cm}
	\ForceWidth{7.5cm}
	\centerline{\BoxedEPSF{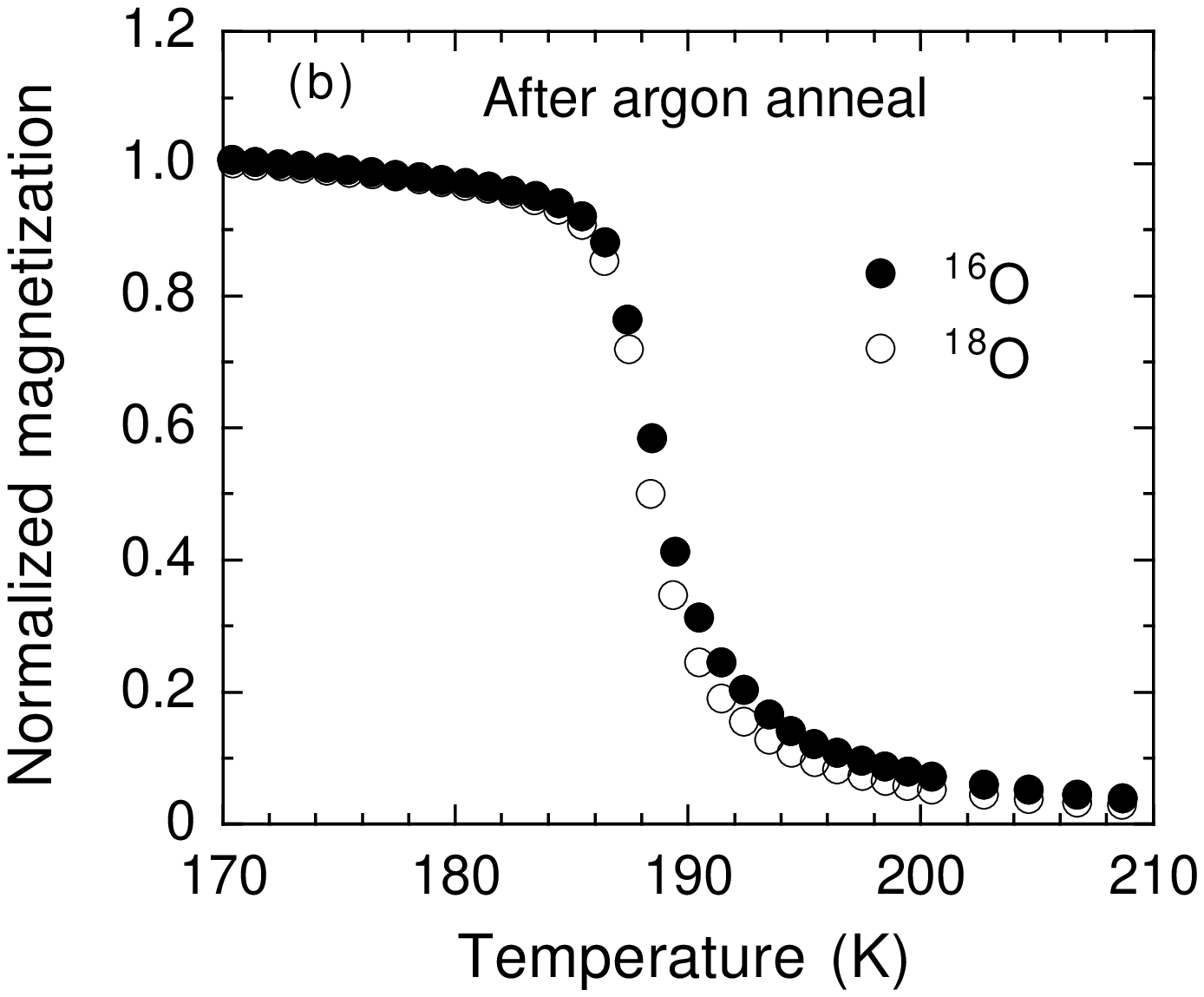}}
	\vspace{0.5cm}
	\caption[~]{Temperature dependence of the low-field magnetization
for the $^{16}$O and $^{18}$O samples
of La$_{0.80}$Ca$_{0.20}$MnO$_{3}$ (a) before the argon annealing; (b)
after the argon annealing at 950 $^{\circ}$C for 24 h. It is clear that
before the argon annealing, the oxygen-isotope shift of T$_{C}$ is about
21 K, while the shift becomes very small (about 1 K) after the annealing.}
	\protect\label{Fig.2}
\end{figure}

It is also important to see how sensitively the T$_{C}$ depends on
the argon annealing. In Fig.~3, we show the temperature dependence of
the low-field magnetization for the $^{16}$O sample before and after 24 hour
argon annealing. The argon annealing does not cause a decrease in
the T$_{C}$ of the $^{16}$O sample. This is in contrast to the result
shown in Ref.~\cite{Franck98} where
the argon annealing leads to a decrease of T$_{C}$ by about 10 K. The
discrepancy is possibly due to the fact that our present $^{16}$O
sample is nearly stoichoimetric. From the TG data shown in Fig.~1, one
can see that the oxygen content of the $^{16}$O sample decreases by
about 0.04 per unit cell after the argon annealing. This implies that
the T$_{C}$ is very insensitive to the oxygen content in the present
sample where the oxygen content is nearly stoichoimetric. The present
result is
consistent with Ref.~\cite{Tamura2} where it was shown that
the T$_{C}$ of the stoichoimetric sample of La$_{0.8}$Ca$_{0.2}$MnO$_{3}$ is
reduced by about 3 K when introducing
about 0.05 oxygen vacancies per cell. This would imply that the
oxygen content of the $^{18}$O sample must be smaller by 0.35 per cell
than the $^{16}$O sample in order to produce the observed oxygen
isotope shift of 21 K. In fact, it was shown that
the difference in the oxygen content of the $^{16}$O and $^{18}$O samples
of (La$_{0.25}$Pr$_{0.75}$)$_{0.7}$Ca$_{0.3}$MnO$_{3}$  is
less than 0.002 per cell, while the isotope shift is
larger than 100 K \cite{Balagurov}. Moreover, our normal
oxygen-isotope exchange procedure has been extensively used for the
isotope effect experiments in cuprates
\cite{ZhaoNature1,ZhaoJPCM1,ShenPRL}. Both indirect
\cite{ZhaoNature1,ZhaoJPCM1} and direct \cite{ShenPRL} measurements on
the oxygen content consistently
show that the oxygen contents of two isotope samples are the same
within 0.0003 per cell. Therefore, the observed
large oxygen-isotope shift cannot be caused by a negligible difference
in the oxygen
stoichiometries of the two isotope samples.
\begin{figure}[htb]
    \ForceWidth{7.5cm}
	\centerline{\BoxedEPSF{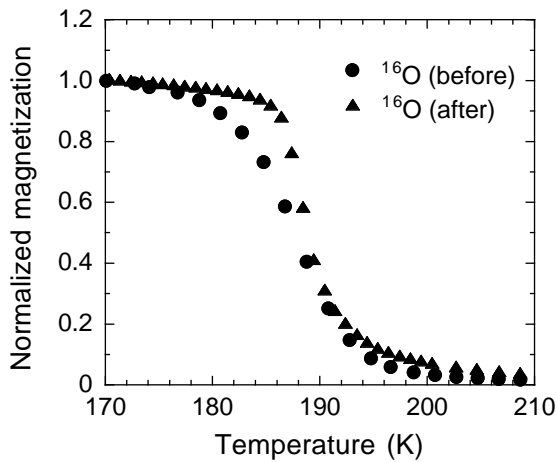}}
	\vspace{0.5cm}
	\caption[~]{Temperature dependence of
the low-field magnetization for the $^{16}$O sample
of La$_{0.80}$Ca$_{0.20}$MnO$_{3}$ before (solid circle) and after
(solid triangle) argon annealing for 24 h. The argon annealing does not
cause a significant
change in the T$_{C}$ of the $^{16}$O sample.}
	\protect\label{Fig.3}
\end{figure}

Our present result, together with some isotope-effect results from
other groups, can strongly argue against the theoretical model
proposed by Nagaev \cite{Nagaev}. According to his model, the $^{16}$O
sample always has more oxygen content than the $^{18}$O sample if the
samples are nonstoichoimetric. The more nonstoichoimetric the samples
are, the more difference in the oxygen contents of two isotope samples,
and thus the larger the isotope effect is. This is in contradiction
with experiment. The stoichoimetric
(La$_{0.25}$Pr$_{0.75}$)$_{0.7}$Ca$_{0.3}$MnO$_{3}$ compound shows a
very large isotope effect \cite{Balagurov}, while the very
nonstoichoimetric (LaMn)$_{0.945}$O$_{3}$ material has a rather small
isotope effect \cite{ZhaoPRL}. As a matter of fact, the isotope
exponent is proportional to the pressure-effect coefficient, and
simply depends on T$_{C}$ \cite{ZhaoPRB}. Furthermore, this
theoretical model would predict a negative oxygen isotope effect
(i.e., the $^{18}$O sample has a higher T$_{C}$) for
the overdoped regime where $dT_{C}/dx$ $<$ 0. In reality, one has always
found positive isotope effects \cite{Franck98,ZhaoPRB}. Thus, we must
conclude that the theoretical explanation to the observed isotope
effects by Nagaev \cite{Nagaev} is not correct.

In summary, our present TG experiments clearly demonstrate that the
argon annealing on the oxygen-isotope
exchanged samples causes the $^{18}$O sample to
partially exchange back to the $^{16}$O isotope due to a small $^{16}$O
contamination in the Ar gas. Such a contamination is commonly caused by
the oxygen outgas that is trapped
in the tubes, connectors and valves. The present result clearly shows
that the oxygen-isotope effect observed in the argon annealed samples may
not be intrinsic, and that the normal isotope exchange procedure can ensure
the same oxygen content for two isotope samples and thus produce an
intrinsic isotope effect.  The observed large oxygen isotope effect
which is an intrinsic property of manganites places an important
constraint on the basic physics of these materials.
~\\
~\\
\noindent
{\bf Acknowlegments}: We would like to thank A. S. Alexandrov for
useful discussion. This work was supported
by the Swiss National Science Foundation.

\bibliographystyle{prsty}


\end{document}